## Euler Analysis of a Machine for Raising Waters Proposed by Mr. de Mour

Sylvio R. Bistafa
sbistafa@usp.br
University of São Paulo, Brazil

**Abstract**

We revisited an analysis made by Euler in a memoir of 1753 *Sur une nouvelle manière d'élever de l'eau propose par M. de Mour* ("On a new method to raise water proposed by Mr. de Mour"), addressing a type of water sprinkler, consisting of an inclined tube with its lower end immersed into water, and discharging water at its top by turning a vertical axis to which the inclined tube is attached.

**Keywords**: earlier machines for raising waters, water sprinkler for irrigation, water pumps

### 1. Introduction

Many publications attest Euler's recurrent interest in machines for raising waters. The theory and application of piston pumps were addressed at least in four publications (E206, E207, E409Ch3, E409Ch4)[a]. By its turn, the theory of hydraulic rotating wheels was addressed in three different publications (E179, E202, E259). These latter works were eventually consolidated in E222, where Euler presents what is considered the pioneering general theory of rotating hydraulic machines.

The present work revisits an analysis made by Euler (1753, E203) of a particular type of rotating wheel to raise waters, proposed by a certain Mr. le Demour (1732)[b], which is described (in French) as follows:

> *It is not really a Machine, so simple it is. It is certain that when an inclined Pipe is moved into Water with a small force, the water will rise to a certain height, & it will exit at the top end, provided the Pipe is not too long. It is also necessary that the lower end be cut like a flute mouthpiece, so that it catches the water better, & picks it up more easily. This is the whole principle, and there is nothing more to do other than a small Frame which holds the pipe inclined, such that in this way, it can be easily moved into the water, which is very easy to imagine & execute. The Machine is put in motion by a unique man driving a crank, making 34 revolutions in certain number of seconds, & raised about 220 pints of water to a height of 6 feet. The pipe was inclined to 50 degrees. This quantity of water is the same as that of the best Frame to a height of 8 feet, by applying 4 men to it. We can see how much the new Machine would save in trouble and expense. The principal intention of the Author is to use it for the irrigation of lands less expensively than ground waters or rivers, to change arable land into meadows, to improve funds, which it is not too much neglected by those who are the masters.*

---

[a] These are known as Eneström index, which are used to identify Euler's writings. Most historical scholars refer to the works of Euler by their Eneström index. They were introduced by Gustaf Hjalmar Eneström (5 September 1852 – 10 June 1923), a Swedish mathematician, statistician and historian of mathematics.
[b] In E203, Euler spells *de Mour* and indicates № 363 as (presumably) the reference page, which, in fact appears on page 168.

Based on this description, in E203 Euler proceeds right on to the analysis of the machine which has been translated from French by the present author.

## 2. Analysis of the Machine[c]

I. The machine (Fig. 1) is composed of an inclined tube $CD$, which is attached to a vertical rotating axis $OO$. The machine is immersed into water up to $vv$, and by turning the tube $CD$ around the axis $OO$, the water will raise through the tube from the inlet $Cc$, discharging at the outlet $Dd$. It is seen that it is possible to attach to the same axis $OO$, several tubes $Cd$ with the same inclination, in order to increase the output of the machine.

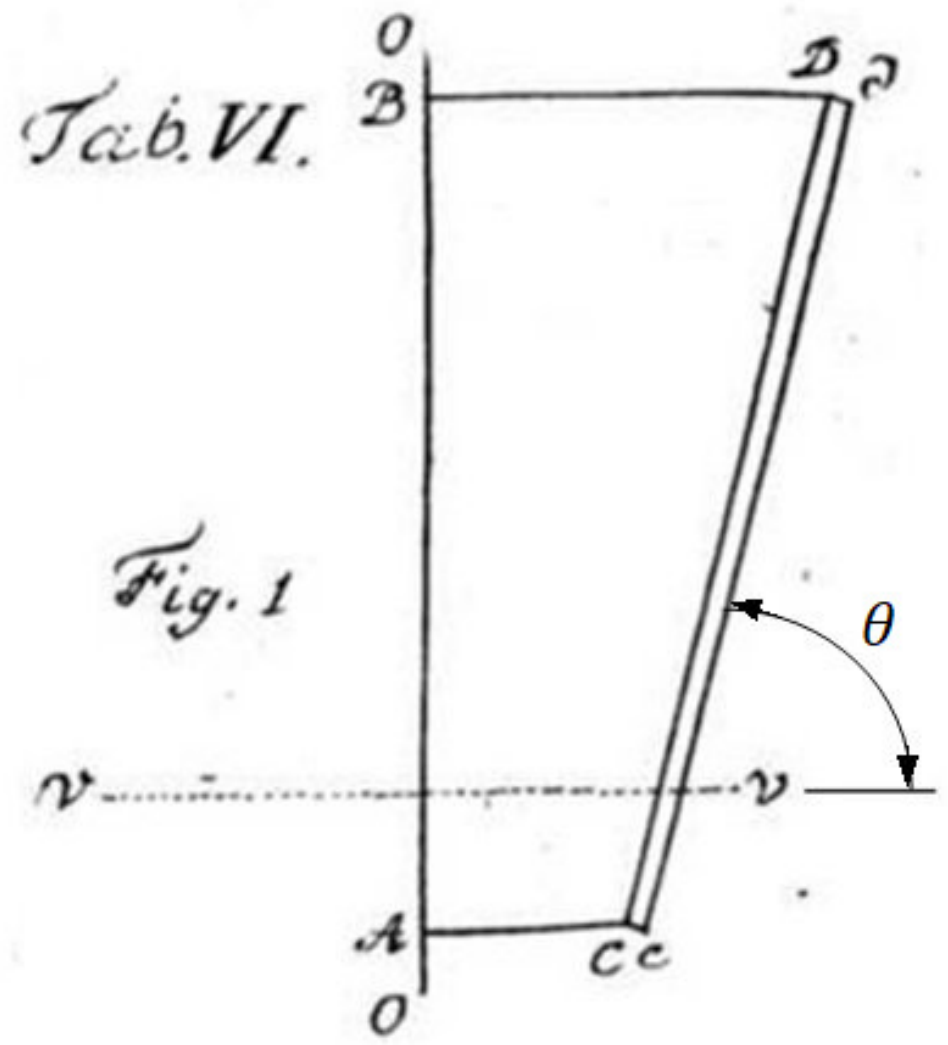

Figure 1: An inclined tube $CD$ attached to a rotating axis $OO$, immersed into water up to $vv$, for raising water through the tube from $Cc$ and discharging at $Dd$. (Adapted from Euler, 1753, Fig. 1)

[Paragraphs II and III were considered of secondary importance and were omitted]

IV. The length of the tube $CD = b$, its amplitude $= bb$, and set the inclination $= \theta$ such that it corresponds to the complement of the inclination of the tube in relation to the axis $OO$. The horizontal segment $AC = c$ and $BD = c + b\cos\theta$, and the height $AB = b\sin\theta$. It is supposed that the tube rotating motion is such that at a distance of the axis $= a$, the velocity of rotation $= \sqrt{k}$ , in which $k$ is the due height, such that the absolute velocity of rotation (the angular velocity) is $= \frac{\sqrt{k}}{a}$.

V. Set $\sqrt{v}$ the velocity of the water inside the tube, at the present instant of time, which is uniform throughout the tube, and that this velocity is the same through the orifice $Dd$ at the top: and, therefore, the amount of the water discharged in $Dd$ will be expressed by $bb\sqrt{v}$. Finally, the depth of the lower aperture $Cc$ below the surface of the water $vv$ will be set $= e$.

VI. This motion is regarded as being known, such that $v$ will a certain function of time $t$, which is supposed has already elapsed since the beginning of the motion; then it is necessary to find the forces, which each particle of the water inside the tube is subjected to, for the supposed motion to take place. Then, when these forces are compared to those which the water is currently solicited, that it will be possible to determine not only the actual motion, but also the force which must be applied to keep the machine in motion.

[c] Some paragraphs containing side information were omitted, some were abridged and some were merged together, to save space and reader's time without compromising the essentials of the developments.

VII. To this effect (Fig. 2), let us decompose the motion according to three mutually perpendicular fixed axis, where $OO$ is the vertical axis, which coincides with the axis around which the machine turns about, and the remaining two axis $OA$ & $OB$ are in the same horizontal plane, over which the radius $OD = c$ describes the circle $DPE$, and after a certain time $= t$, when the velocity of the water through the tube is supposed $= \sqrt{v}$, the tube is at the position $PQ$, having from the beginning already described by its rotating motion the angle $D\hat{O}P = \varphi$. Therefore, since the velocity with which the point $P$ is carried towards $E$ is $= \frac{c\sqrt{k}}{a}$, describing in the infinitesimal time $dt$ the arc $= cd\varphi$, we will have that $d\varphi = \frac{dt\sqrt{k}}{a}$, and, hence, $\varphi = \frac{t\sqrt{k}}{a}$, or $t = \frac{a\varphi}{\sqrt{k}}$, such that the angle $D\hat{O}P = \varphi$ will serve as a measure of time.

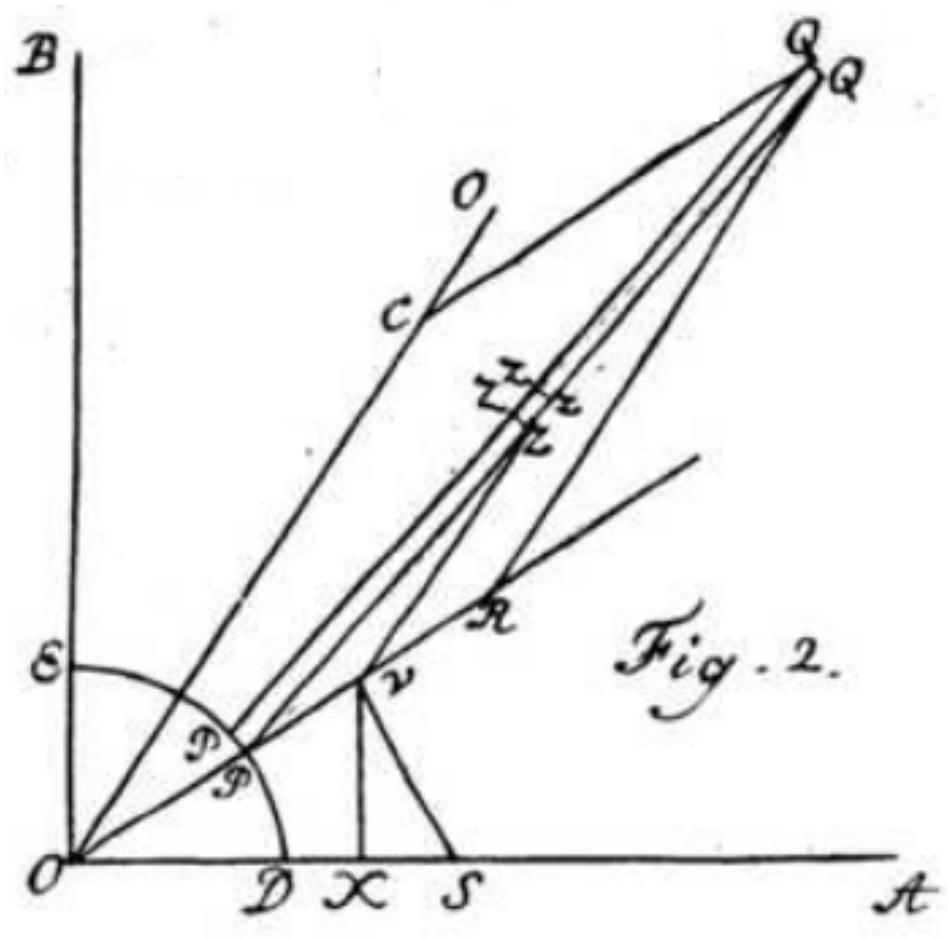

Figure 2: Tube projection $PQ$ in the horizontal plane $OAB$. (Euler, 1753, Fig. 2)

VIII. Let us consider an element of water $Zz$ inside the tube at a distance $PZ = s$, and by putting $Zz = ds$, then, the mass of water of this element of water will be $= b^2 ds$. Let us drop from the point $Z$, the vertical $ZY$, and, because the angle $Y\hat{P}Z = \theta$, we will have that $Zz = ds = dt\sqrt{v}$. Additionally, let us draw from the point $Y$, the perpendicular $YX$ to the axis $OA$, and, because the angle $A\hat{O}Y = \varphi$, we will have that $OX = (c + s\cos\theta)\cos\varphi$, & $XY = (c + s\cos\theta)\sin\varphi$.

IX. Calling $OX = x, XY = y$ & $YZ = z$, then, $x = (c + s\cos\theta)\cos\varphi$, $y = (c + s\cos\theta)\sin\varphi$ & $z = s\sin\theta$; and, considering that the element $dt$ is constant, by the principles of Mechanics it is then necessary that the particle of water in $Z$ be solicited by three accelerating forces, which are

According to $OX$:

$$\frac{2d^2x}{dt^2} = \frac{2}{dt^2}(-cd^2\varphi\sin\varphi - cd\varphi^2\cos\varphi + d^2s\cos\theta\cos\varphi - ds\,d\varphi\cos\theta\sin\varphi - sd^2\varphi\cos\theta\sin\varphi - sd\varphi^2\cos\theta\cos\varphi).$$

According to $XY$:

$$\frac{2d^2y}{dt^2} = \frac{2}{dt^2}(-cd^2\varphi \cos\varphi - cd\varphi^2 \sin\varphi + d^2s \cos\theta \sin\varphi + ds\, d\varphi \cos\theta \cos\varphi + sd^2\varphi \cos\theta \cos\varphi - sd\varphi^2 \cos\theta \sin\varphi).$$

According to $YZ$:

$$\frac{2d^2z}{dt^2} = \frac{2}{dt^2}(d^2s \sin\theta).$$

X. By taking $YS$ perpendicular to $OY$, then the two other forces according to $OY$ and $YS$ can be written, respectively, as $OX \cos\varphi + XY \sin\varphi$ and $OX \sin\varphi - XY \cos\varphi$. Consequently, the element of water in $Z$ is subjected to the following three accelerating forces

I. According to $OY$:

$$\frac{2}{dt^2}(-cd\varphi^2 + d^2s \cos\theta - sd\varphi^2 \cos\theta).$$

II. According to $YS$:

$$\frac{2}{dt^2}(-cd^2\varphi - 2ds\, d\varphi \cos\theta - sd^2\varphi \cos\theta).$$

III. According to $YZ$:

$$\frac{2}{dt^2} d^2s \sin\theta.$$

XI. Knowing that $\frac{d\varphi}{dt} = \frac{\sqrt{k}}{a}$, and $\frac{ds}{dt} = \sqrt{v}$, then, $\frac{d^2\varphi}{dt^2} = 0$, $\frac{d^2s}{dt^2} = \frac{dv}{2dt\sqrt{v}}$, $\frac{d\varphi^2}{dt^2} = \frac{k}{a^2}$, and $\frac{dsd\varphi}{dt^2} = \frac{\sqrt{kv}}{a}$: therefore, the three accelerating forces can be rewritten as

I. According to $OY$:

$$-\frac{2ck}{a^2} + \frac{dv \cos\theta}{dt\sqrt{v}} - \frac{2ks \cos\theta}{a^2}.$$

II. According to $YS$:

$$-\frac{4\cos\theta\sqrt{kv}}{a}.$$

III. According to $YZ$:

$$\frac{dv \sin\theta}{dt\sqrt{v}}.$$

XII. In case these forces don't act in each particle of the water, they would pursue their motion by virtue of their inertia: it is therefore the impenetrability of the walls of the tube which provides these forces; and consequently, these walls will be reciprocally solicited by equal forces in the opposite directions. But, of these three forces there is only the one which acts according to $YS$, which opposes the rotational motion of the machine. This driving force acting on the element $Zz = b^2ds$ will be, therefore, $= \frac{4b^2 ds\, cos\, \theta\sqrt{kv}}{a}$, and the moment will be $= \frac{4b^2}{a} ds(c + s\, cos\, \theta)\, cos\, \theta\sqrt{kv}$. And, for all the water contained inside the tube, gives the following moment that will oppose the motion of the machine

$$\frac{2b^2}{a}(2bc + b^2\, cos\, \theta) \cos\theta\sqrt{kv} = \frac{2b^3 cos\, \theta}{a}(2c + b\, cos\, \theta)\, \sqrt{kv}.$$

XIII. Then, the water particle in $Z$ will be pushed according to the direction up the pipe, or according to $ZQ$, by the accelerating force $= OY \cos\theta + YZ \sin\theta$, and, consequently, it will be $= \frac{dv}{dt\sqrt{v}} - \frac{2ckcos\, \theta}{a^2} - \frac{2ks\, cos^2\theta}{a^2}$. It is then necessary that the water be actually solicited by this force according to the direction $ZQ$ of the tube.

XIV. But, actually, there are two forces acting on the water in $Z$, the first is the gravity, but since it acts according to $ZY$, there is no moment according to the rotating motion of the machine, however, the water in $Z$ will be solicited according to $ZP$ with the accelerating force [per unit of weight] $= sin\, \theta$. The other force is due to the state of compression of the water inside the tube. Be the pressure on $Z$ expressed by the height $= p$, and on $z$ by $p + dp$: and, therefore, each particle of the water in $Zz$ will be driven to the lower end of the tube according to $ZP$ by the accelerating force $= \frac{dp}{ds}$. By adding together these two forces, we will have the following equation

$$\frac{dv}{dt\sqrt{v}} - \frac{2ckcos\, \theta}{a^2} - \frac{2ks\, cos^2\theta}{a^2} = -\, sin\, \theta - \frac{dp}{ds}$$

or

$$dp = -\frac{dv}{dt\sqrt{v}} ds + \frac{2ckdscos\, \theta}{a^2} + \frac{2ksds\, cos^2\theta}{a^2} - ds\, sin\, \theta.$$

XV. Since $v$ is a function of the time $t$, and because we are looking for the state of compression in each section of the tube at the present time, we should regard the expression $\frac{dv}{dt\sqrt{v}}$ as constant, and, after integration we find

$$p = C - \frac{s\, dv}{dt\sqrt{v}} + \frac{2ckscos\, \theta}{a^2} + \frac{ks^2\, cos^2\theta}{a^2} - s\, sin\, \theta.$$

Since at the exit in $Q$, the pressure should vanish; this circumstance can be used to determine the constant $C$, that, then, will be given by

$$C = \frac{b\, dv}{dt\sqrt{v}} - \frac{2bckcos\, \theta}{a^2} + \frac{b^2 k\, cos^2\theta}{a^2} + b\, sin\, \theta\, ,$$

which gives the state of compression of the water at the lower end $P$.

XVI. Since the orifice is at a depth $= e$ below the surface of the water, through which the water is supposed to be admitted with a velocity $= \sqrt{v}$, then the pressure[d] there should be $= e - v$, giving the following equation

[d] This is the pressure head at the inlet orifice, given by the static head $e$ minus the kinetic head $v$.

$$e - v = \frac{b\,dv}{dt\sqrt{v}} - \frac{2bckcos\,\theta}{a^2} + \frac{b^2k\,cos^2\theta}{a^2} + b\,sin\,\theta$$

and, supposing that the steady state condition has been reached, with a uniform velocity $\sqrt{v}$ throughout the tube, then we will have the following equation

$$v = e + \frac{bk\,cos\,\theta}{a^2}(2c + b\,cos\,\theta) - b\,sin\,\theta,$$

where $b\,sin\,\theta - e$ is the height of the upper orifice of the tube in relation to the free surface of the water $vv$.

XVII. For the water to raise through the tube, it is necessary that the velocity of rotation surpasses a certain limit, or it is necessary that

$$\frac{k}{a^2} > \frac{b\,sin\,\theta - e}{b\,cos\,\theta\,(2c + b\,cos\,\theta)}.$$

XVIII. Let us suppose that the rotating motion is such that each complete revolution of the machine corresponds to the oscillations of a pendulum of length $= q$, then, $\frac{k}{a^2} = \frac{2}{q}$[e], and, therefore,

$$v = \frac{2b\,cos\,\theta}{q}(2c + b\,cos\,\theta) - b\,sin\,\theta + e;$$

from which

$$q < \frac{2\,b\,cos\,\theta\,(2c + b\,cos\,\theta)}{b\,sin\,\theta - e},$$

and the moment of the forces which opposes the motion of the machine is

$$= \frac{2b^3cos\,\theta}{a}(2c + b\,cos\,\theta)\sqrt{\frac{2v}{q}}.$$

XIX. The machine can be easily built from the limit given above for the length $q$ of the pendulum, as follows. Set $AE = AC = c$ (Fig.3), and let us drop from $D$ on the horizontal line $EAC$, the perpendicular $DF$. Let us draw the perpendicular $GCH$ to $DE$ passing through $C$, until it meets the prolonged vertical$DF$; and suppose that the segment $EF$ is the level of the water, such that $e = 0$, then it is necessary that $q < 2FH$[f].

XX. But, these findings still require a correction, regarding the water pressure in $C$, which we have supposed $= e - v$, which would be correct if the orifice of the tube $C$ were at rest, which would happen if it were $c = 0$; such that for the case where $c = 0$, our formulas would not need correction. But if the orifice $C$ turns around the axis $OO$ with the velocity $= \frac{c\sqrt{k}}{a}$, it would be the same if the tube were at rest, while the water turns with an equal velocity around the axis $OO$. Or in this case, the water pressure in $C$ would be greater because of the centrifugal force, so that in place of the height $e$ it is necessary to write $e + \frac{c^2k}{a^2}$; and, hence, I should have written $e + \frac{c^2k}{a^2} - v$ instead of $e - v$ in the equation obtained in § 16.

---

[e] The distance covered in a complete revolution the machine with the velocity $\sqrt{2gk}$ is $2a\pi = \sqrt{2gk} \cdot \frac{T}{2}$, where $\frac{T}{2} = \pi\sqrt{\frac{q}{g}}$, is the time for one swing of the pendulum, then, $\frac{k}{a^2} = \frac{2}{q}$.

[f] We have that $(2c + b\,cos\,\theta) = EF$, and also that $tan\,\beta = \frac{EF}{DF} = \frac{EF}{b\,sin\,\theta} = \frac{FH}{b\,cos\,\theta}$, then $FH = \frac{b\,cos\,\theta\,EF}{b\,sin\,\theta} = \frac{b\,cos\,\theta(2c+b\,cos\,\theta)}{b\,sin\,\theta}$, for $e = 0$.

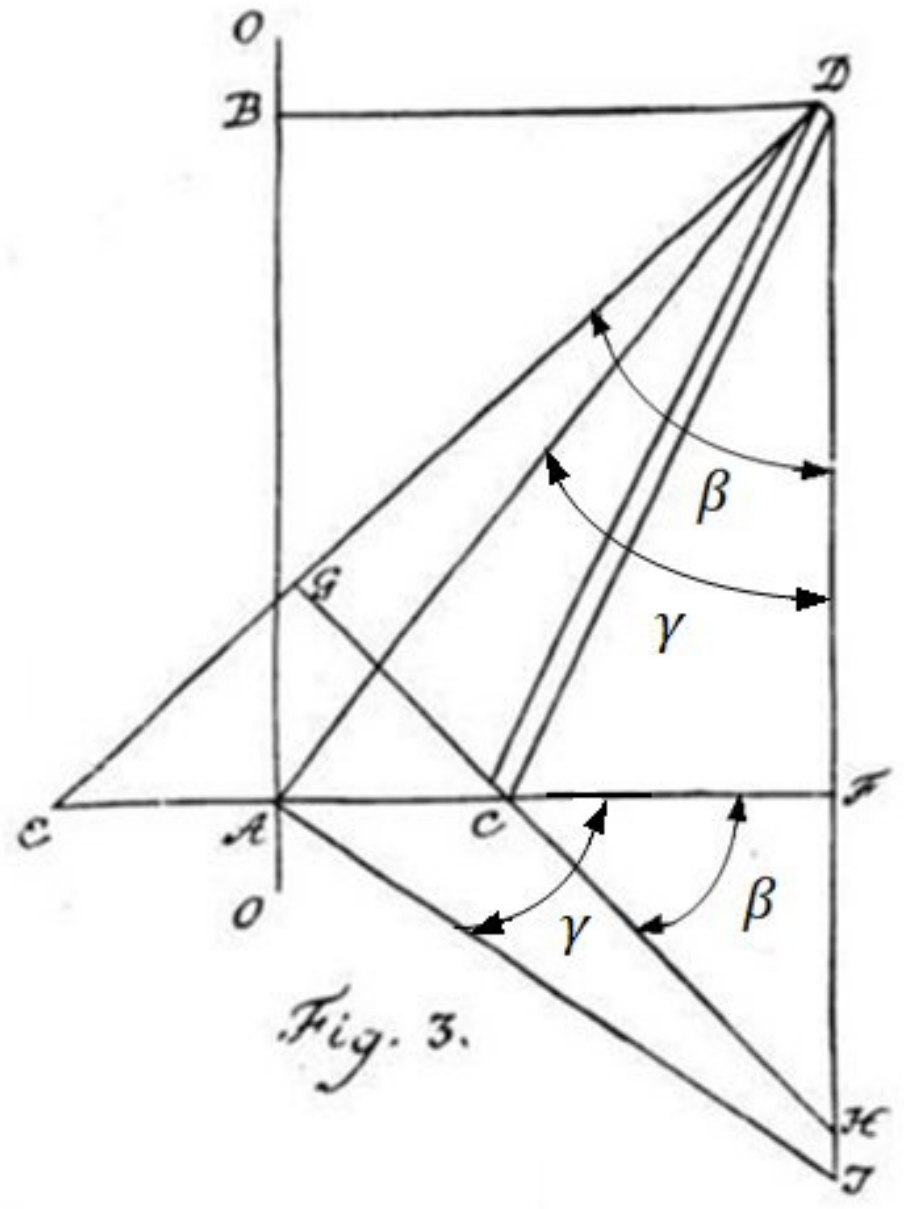

Figure 3: Adapted from Euler, 1753, Fig. 3.

XXI. Since $\frac{k}{a^2} = \frac{2}{q}$, if we put $e + \frac{2c^2}{q}$ in place of $e$, we will have the following equation

$$v = \frac{2}{q}(c + b\cos\theta)^2 - b\sin\theta + e,$$

then, if we consider that the height $e$ is infinitely small, we will have that

$$v = \frac{2}{q}(c + b\cos\theta)^2 - b\sin\theta.$$

Therefore, for the water to actually raise through the tube it s necessary that

$$q < \frac{2\,(c + b\cos\theta)^2}{b\sin\theta}.$$

To construct the value of this formula, one has to draw from the segment $DA$, the perpendicular $AJ$ which cuts the vertical $DF$ in $J$, and, therefore, we have that $q < 2FJ$.[g]

[g] We have that $(c + b\cos\theta) = AF$, and also that $\tan\gamma = \frac{AF}{DF} = \frac{AF}{b\sin\theta} = \frac{FJ}{c+b\cos\theta}$, then $FJ = \frac{(c+b\cos\theta)\,AF}{b\sin\theta} = \frac{(c+b\cos\theta)^2}{b\sin\theta}$.

XXII. If we want the water to raise through the tube with a velocity $\sqrt{v}$, it is possible to determine the rotating velocity necessary to turn the machine: since each revolution should be achieved during the time of one oscillation of a pendulum, whose length is

$$q = \frac{2(c + b\cos\theta)^2}{v + b\sin\theta};\ or, \frac{2}{q} = \frac{v + b\sin\theta}{(c + b\cos\theta)^2}.$$

So that the moment of the forces necessary to keep the machine in motion will be

$$= \frac{2b^3\cos\theta(2c + b\cos\theta)}{(c + b\cos\theta)}\sqrt{v(v + b\sin\theta)}.$$

Therefore, it will be necessary to use these formulas instead of the previous ones if the pipe is extended downwards to the axis $O$, to draw the water from there.

XXIII. Now, unless $c = 0$, the water which enters the orifice at $C$ will also exert some force contrary to the motion of the machine, and considering this force, the total moment of forces which opposes the motion of the machine will be $= \frac{2b^2}{a}(c + b\cos\theta)^2\sqrt{kv}$. Hence, since $\frac{\sqrt{k}}{a} = \sqrt{\frac{2}{q}} = \sqrt{\frac{v(v+b\sin\theta)}{(c+b\cos\theta)^2}}$, then, this moment will be $= 2b^2(c + b\cos\theta)\sqrt{v(v + b\sin\theta)}$.

XXIV. But, to consider once and for all the foundation of the corrections, which it is necessary to bring with regard to the distance $AC = c$, where the lower orifice draws water; the best way will be to consider a curved tube according to any figure, and to give it a variable width. For this generality will serve to make us see the difference that there is, whether the distance $AC = c$ is vanishing, or not; since we can always conceive the shape of the pipe such that its socket opening $C$ taps near the same axis, so that there is no longer any difficulty on the side of the distance $c$.

XXV. Let us now consider that the tube $PQ$ (Fig. 4) has any figure, which is, however, in the same plane with the axis $OO$, and setting the portion $PZ = s$, the width of the tube in $Z = r^2$, and, as before, the width at the end $QQ = b^2$, where the water exits with a velocity $= \sqrt{v}$. Hence, the velocity of the water in $Z$ will be $= \frac{b^2\sqrt{v}}{r^2}$, with which it covers during the element of time $dt$ the distance $ds = \frac{b^2 dt\sqrt{v}}{r^2}$, where it is necessary to point out that $r^2$ is a function of the distance $s$ and independent of the time $t$, whose quantity $v$ is a function.

[Paragraphs XXVI, XXVII, XXVIII were condensed into one paragraph]. Let us set $ZY = z$ and $OY = u$, which will be the coordinates for the figure of the tube, on which depend the ratio between the quantities $s, z$ and $u$. So be $dz = ds\sin\omega$ and $du = ds\cos\omega$, or $dz = \frac{b^2 dt\sin\omega\sqrt{v}}{r^2}$, and $du = \frac{b^2 dt\cos\omega\sqrt{v}}{r^2}$. Next, let us set $OX = u\cos\varphi = x$ and $YX = u\sin\varphi = y$, and the three accelerating forces which act on the water in $Z$ will be: according to OX $= \frac{2d^2x}{dt^2}$; according to XY $= \frac{2d^2y}{dt^2}$ and according to YZ $= \frac{2d^2z}{dt^2}$.

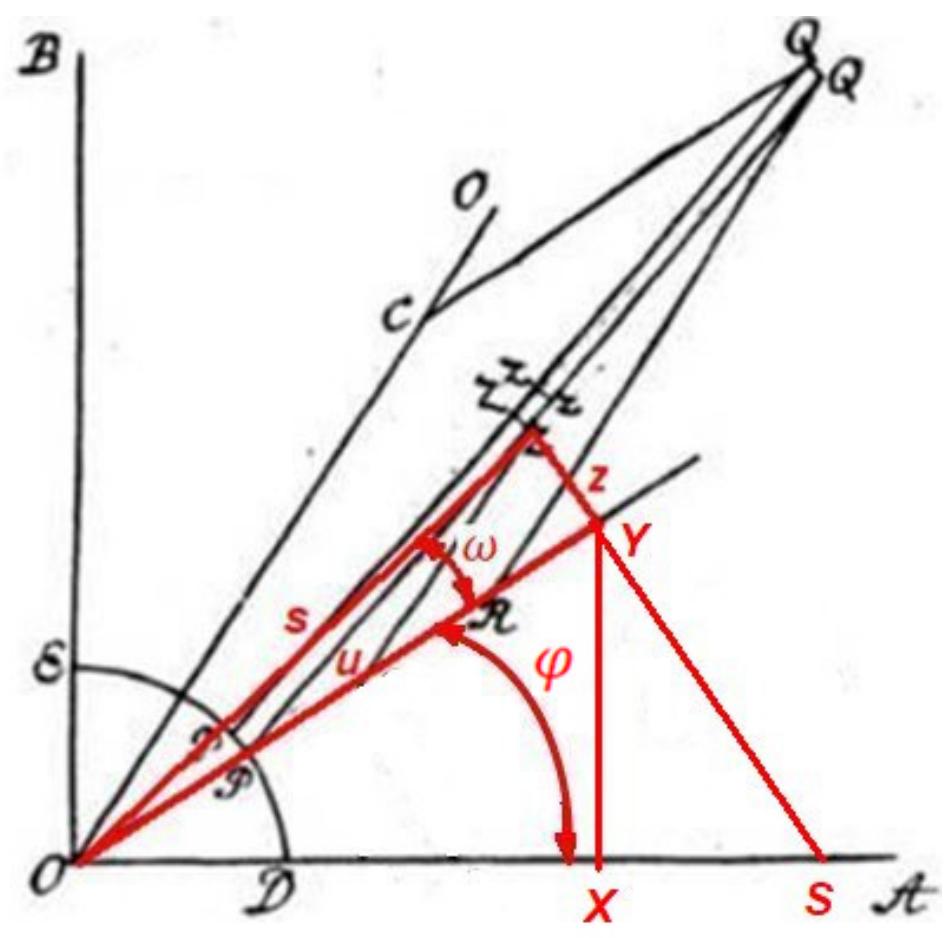

Figure 4: Adapted from Euler, 1753, Fig. 2.

From the decomposition of the forces according to $OX$ and $XY$, Euler finds that the only force that opposes the rotational motion of the machine is the force that acts according to $YS$, which would be given by $-\frac{4b^2 cos\,\omega\,\sqrt{kv}}{ar^2}$.

XXIX. Then, the resulting element of the driving force will be $= -\frac{4b^2 ds\, cos\,\omega\,\sqrt{kv}}{a}$[h], and the element of the moment will be $= \frac{4b^2 u\, du\,\sqrt{kv}}{a}$, giving the total moment as $\frac{2b^2 u^2\,\sqrt{kv}}{a}$. Hence, if the tube begins at $P$ and ends at $Q$, and by dropping the perpendicular from $Q$, the total moment which results from all the water contained inside the tube is

$$\frac{2b^2\sqrt{kv}}{a}(OR^2 - OP^2).$$

XXX. The water particle in $Z$ will be pushed according to the direction up the pipe, or according to $ZQ$, by the accelerating force $= OY\,cos\,\omega + YZ\,sin\,\omega$,in which $OY$ is obtained from the decomposition of the forces according to OX and XY, resulting in

$$-\frac{4b^2 dr\sqrt{v}}{r^3 dt} + \frac{b^2 dv}{r^2 dt\,\sqrt{v}} - \frac{2ku\,cos\,\omega}{a^2},$$

which should be $= -\frac{dp}{ds} - sin\,\theta$, supposing that the pressure in $Z$ is equal to $p$.

From these, Euler obtains

$$p = C - z - \frac{b^4 v}{r^4} + \frac{ku}{a^2} - \frac{b^2 dv}{dt\,\sqrt{v}}\int\frac{ds}{r^2}\,.$$

XXXI. By calling for the whole length of the tube $\int\frac{ds}{r^2} = F$, and since the pressure in $Q$ should vanish, where $z = QR, u = OR$ and $r^2 = b^2$, then

[h] In fact, $-\frac{4b^2 cos\,\omega\,\sqrt{kv}}{ar^2}$ is the acceleration, which when multiplied by the volume per unit of weight $r^2 ds$, gives the force per unit of weight $-\frac{4b^2 ds\, cos\,\omega\,\sqrt{kv}}{a}$.

$$C = QR + v - \frac{k}{a^2} \cdot OR^2 + \frac{b^2 dv}{dt\,\sqrt{v}} \cdot F.$$

Supposing that the width of the tube in $P$ is equal to $f^2$, then the pressure in $P$ will be

$$\frac{F\,b^2 dv}{dt\,\sqrt{v}} + QR - \frac{k}{a^2}(OR^2 - OP^2) + v\left(1 - \frac{b^4}{f^4}\right).$$

XXXII. Now, supposing that $OP = 0$, or that the tube of any figure has at its lower end a curvature such that the orifice there is next to the axis, there is no doubt that the pressure will be $= e - \frac{b^4 v}{f^4}$, giving $e = \frac{F\,b^2 dv}{dt\,\sqrt{v}} + QR - \frac{k}{a^2} OR^2 + v$, and if the motion is already in its steady state, then $v = \frac{k}{a^2} OR^2 - QR + e$, and the moment of the forces $= \frac{2b^2\sqrt{kv}}{a} OR^2$.

XXXIII. It is seen that to have the greatest possible velocity, it is advantageous to keep the lower orifice next to the axis, because in this case, the negative contribution of $OP^2$ to the velocity would vanish.

XXXIV. So be $OO$ the axis of the machine, and $AMD$ the figure of the tube, whose lower orifice $Cc$ draws water very close to the axis $OO$, and that the upper opening $Dd$ is bent outwards to better disgorging the water; whose opening is $= b^2$. As far as the width of the tube in the other places $Mm$, it does not matter whether it is larger or smaller than $b^2$; but to decrease the hindrance of friction, and to allow the motion of the water more freely, it will be advisable to give to the tube at the bottom a greater width.

XXXV. This tube being, therefore, turned around the axis so quickly that each revolution responds to an oscillation of a pendulum of the length $= q$, and since $\frac{k}{a^2} = \frac{2}{q}$, the water will raise through the tube with the velocity $\sqrt{v}$, and, then, $v = \frac{2 \cdot EF^2}{q} - AE$, assuming that $AE$ is the height above the water level, and to keep the machine in motion, it will be necessary to employ a force whose moment is $= 2b^2 \cdot EF^2 \sqrt{\frac{2v}{q}}$, where the force is expressed by a volume of water whose weight is equal to this volume.

XXXVI. So, if the time of one revolution of the machine is given by the pendulum $q$, it is necessary that $EF^2 > \frac{1}{2} q \cdot AE$ [from § XXI, $q < \frac{2\,(c + b\cos\theta)^2}{b \sin\theta}$].

XXXVII. Therefore, we understand that the most appropriate figure that we can give to the tube $AMF$ is the parabola, whose axis is the axis of the machine, and the top of it is at $A$. So be a parabola the figure of the tube $AMF$, whose parameter is $= g$, such that $PM^2 = g \cdot AP$ and $EF^2 = g \cdot AE$, and the velocity of the water that will be discharged above is due to the height $v = \left(\frac{2g}{q} - 1\right) AE$, and to keep the machine in motion, it is necessary to apply a force whose moment is $= \frac{2b^2}{q} \cdot EF^3 \sqrt{\left(4 - \frac{2q}{g}\right)}$, or this moment will be $= \frac{2gb^2}{q} \cdot AE\sqrt{AE(4g - 2q)}$.

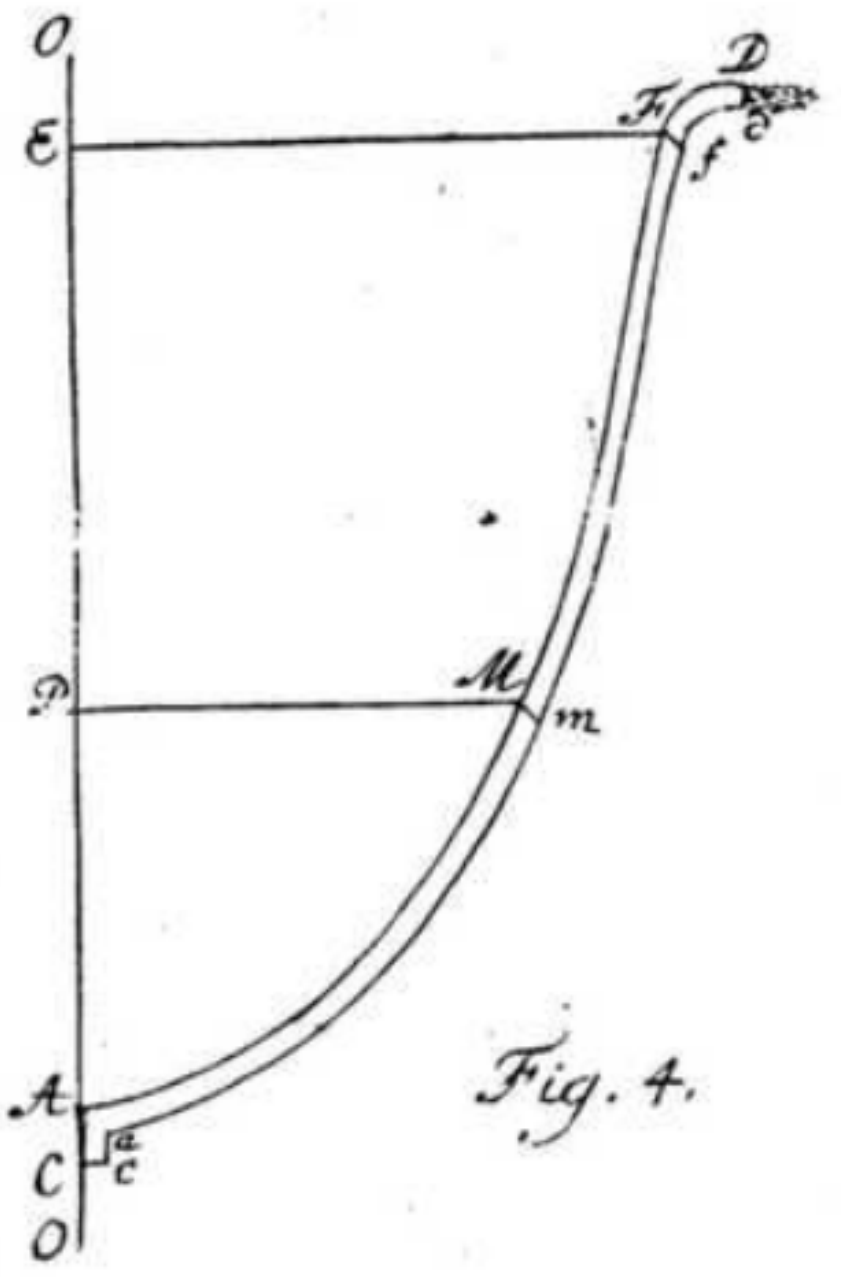

Figure 5: Parabolic tube. (Euler, 1753, Fig. 4)

XXXVIII. Then, it is necessary that the parabola parameter $g$ be greater than $\frac{1}{2}q$ [$EF^2 > \frac{1}{2}q \cdot AE => g \cdot AE > \frac{1}{2}q \cdot AE => g > \frac{1}{2}q$]. Let us set $\frac{2g}{q} - 1 = \propto$ [$g > \frac{1}{2}q => \frac{2g}{q} > 1 => \frac{2g}{q} - 1 > 0, then \propto > 0$ ]], giving the parabola parameter as $g = \frac{1}{2}(\propto +1)q$; and the height to which the water should be raised $AE = a$, we have that $v = \propto a$, and the required moment to put the machine in action $= (\propto +1)\, b^2 a \sqrt{2 \propto aq}$.

XXXIX. Set $l$ as the height for which a weight falls in one second 15.625 $ft$; and if it were $v = l$, the mass of water discharged in one second would be $= 2b^2 l$; and since $v = \propto a$, the volume of water raised in one second will be $= 2b^2\sqrt{\propto al}$.[i]

XL. Since the moment $M = (\propto +1)\, b^2 a \sqrt{2 \propto aq}$, then, we have that $b^2 = \frac{M}{(\propto+1)a\sqrt{2\propto aq}}$: and, therefore, the quantity of water raised in one second will be $\frac{2M}{(\propto+1)a}\sqrt{\frac{l}{2g}}$: it is seen that the faster one moves the machine, the larger it will become the quantity of water that one will be able to lift. Next, it can also be seen that to increase the effect of the machine, it is convenient to make the value of $\propto$ as small as possible.

XLI. But, we also easily understand that these decreases cannot go to infinity: because by decreasing $\propto$ and $q$, the upper orifice of the tube $b^2$ would be increased, and since the lower one [$Cc$] cannot be smaller [larger???], its extent would move too far away from the axis [because in § XXXIV Euler recommends to give to the tube at the bottom a greater width than at the top], so that the supposition made in this respect could no longer take place. For this reason, it is necessary to give a reasonable value to the parameter of the parabola $g$, such that $Cc$ be much more smaller than $EF$[j]: and because $Cc$ is at least equal to $b$: it is necessary that $EF^2$ or $\frac{1}{2}(\propto +1)aq$ $[EF^2 = \frac{1}{2}(\propto +1)q \cdot AE = \frac{1}{2}(\propto +1)aq]$ considerably surpasses the value of $b^2 = \frac{M}{(\propto+1)a\sqrt{2\propto aq}}$, or that $\frac{1}{2}(\propto +1)^2 a^2 q\sqrt{2 \propto aq} > M$, and much greater than this.

[i] Since gravity is $= 2l$, then, the water velocity is actually $\sqrt{2 \cdot 2l \cdot \propto a} = 2\sqrt{\propto al}$, and the flow-rate $= 2b^2\sqrt{\propto al}$.

XL [This paragraph number should have been XLII.] . Let us set $\frac{1}{2}(\propto +1)^2 a^2 q\sqrt{2 \propto aq} = \beta M$, such that $\beta$ is a number much greater than one, and we will have that $\sqrt{q} = \sqrt[3]{\frac{2\beta M}{(\propto+1)^2 a^2 \sqrt{2\propto a}}}$; and the quantity of water raised in one second $\frac{2M}{(\propto+1)a}\sqrt{\frac{l}{2g}} = \sqrt[6]{\frac{2^5 \propto l^3 M^4}{(\propto+1)^2 \beta^2 a}}$ [Euler instead writes $\sqrt[6]{\frac{4\propto l^3 M^4}{(\propto+1)^2 \beta^2 a}}$], where it is seen that this quantity will be the greatest if $\propto = 1$.

XLIII. Therefore, set $\propto = 1$, and, then, we have that $g = q$: $\sqrt{q} = \sqrt[3]{\frac{\beta M}{2a^2\sqrt{2a}}}$ or $q = \sqrt[3]{\frac{\beta^2 M^2}{8a^5}}$; $b^2 = \sqrt[6]{\frac{M^2}{\beta^2 a}}$ ; and the quantity of water raised in one second $\sqrt[6]{\frac{2^3 l^3 M^4}{\beta^2 a}}$ [Euler instead writes $\sqrt[6]{\frac{l^3 M^4}{\beta^2 a}}$]: and it should be pointed out that $\frac{b^2}{q} = \frac{a}{\beta}$, or $q = g = \frac{\beta\, b^2}{a}$.

XLIV. Let be considered that several men should be employed to put the machine in action: since the force of a man depends on the applied velocity, let us suppose that in one second he covers the distance $= i$, and that at this velocity, he employs a force equal to the weight of a volume of water $= A$, and that the work this man is used to turn a wheel whose axis is at a distance $= d$; such that the moment of the force will be $= Ad$. Additionally, let us suppose that while he makes one revolution of this wheel, the machine makes $\mu$ revolutions.

XLV. This man will make one revolution in $\frac{2\pi d}{i}$ seconds, and, therefore, the time for one revolution of the machine will be $= \frac{2\pi d}{\mu i}$ seconds, which corresponds to the time of one oscillation of a pendulum whose length is $q$; be $k$ the length of a second pendulum, and, then we will have that $\frac{2\pi d}{\mu i} = \sqrt{\frac{q}{k}}$. Or since $l$ is the height of fall in one second, we will have that[k] $k = \frac{2l}{\pi^2}$, and, therefore, $\frac{2d}{\mu i} = \sqrt{\frac{q}{2l}}$, which gives $q = \frac{8d^2 l}{\mu^2 i^2}$.

XLVI. Therefore, the moment of the force that a man exerts instantly on the machine will be $= \frac{1}{\mu}Ad$; since the machine is suppose to make $\mu$ turns meanwhile the man makes one. And, therefore, we will have that $M = \frac{1}{\mu}Ad$; then, $q = \sqrt[3]{\frac{\beta^2 d^2 A^2}{8\mu^2 a^5}} = \frac{8d^2 l}{\mu^2 i^2}$; which gives $d^4 = \frac{\beta^2 \mu^4 i^6 A^2}{8^4 a^5 l^3}$. Finally, the quantity of water raised in one second will be $\sqrt[6]{\frac{2^3 d^4 l^3 A^4}{\beta^2 \mu^4 a}} = \frac{iA}{2\sqrt{2}a}$ [Euler instead writes $\sqrt[6]{\frac{d^4 l^3 A^4}{\beta^2 \mu^4 a}} = \frac{iA}{4a}$]; where it is seen that this quantity is independent of $\beta$ and $\mu$, and only depend on $A$, $i$ and $a$.

[j] It is not clear why $Cc$ should be much more smaller than $EF$.

[k] The time of one oscillation of a second a pendulum is half of its period, and given by: $1\ second = \pi\sqrt{\frac{k}{2l}}$, where $2l = gravity$, and, then, $k = \frac{2l}{\pi^2}$.

XLVII. Or, to determine the machine itself, knowing the force $A$ together with the distance $i$ that the man covers in one second, and the height $a$ to which the water should be raised, we will have that $d^2 = \frac{\beta\mu^2 A i^3}{64a^2 l\sqrt{al}}$ and $d = \frac{\mu i}{8a}\sqrt{\frac{\beta A i}{l\sqrt{al}}}$; from which one finds $q = g = \frac{\beta A i}{8a^2\sqrt{al}}$ and $b^2 = \frac{Ai}{8a\sqrt{al}}$: where $l = 15.625\, ft$.

XLVIII. By calling $D$ the flow rate provided by the machine, it is seen that $Da = \frac{1}{4}Ai$ expresses the machine effect in one second, and $Ai$ indicates the action that a man exerts during the same time. From which the machine effect is seen to be equal to the fourth part of the action that is actually employed for raising the water.

[Paragraphs XLVIX and L contain side information and were omitted.]

[Paragraph LI considers a man capable of providing a speed of 2 $ft$/s and a force of 30 $livres$ (about 32 $pounds$)]

Since half a cubic feet of water weight about 30 livres, then, $A = \frac{1}{2}$, and in face that $i = 2$, and $l = 15.625\, ft$, we will have that

$$D = \frac{50}{101a};\ b^2 = \frac{\sqrt{4000}}{a\sqrt{a}};q = 31{,}25 \cdot \frac{d^2}{\mu^2};g = \frac{101}{200}\cdot q,$$

and since $EF = ga = \frac{505}{32}\cdot\frac{ad^2}{\mu^2}$, it is required that $\frac{d^2}{\mu^2}$ be several times greater than $\frac{32\sqrt{4000}}{505a^2\sqrt{a}}$ or that $\frac{4}{a^2\sqrt{a}}$.

LII. [Considering that $\mu = 1$ and $d = 1/2$ the parabola parameter will $q = 31{,}25 \cdot \frac{d^2}{\mu^2} = 31{,}25 \cdot \frac{0.5^2}{1^2} = 7.8125$; $g = \frac{101}{200}\cdot 7.8125 = 3{,}9453125\, ft$, which gives a quite convenient machine. And for a sufficient smaller $d$, one will find $g = 3\, ft$.]

[Paragraphs LIII and LIV contain side information and were omitted.]

LV. However, to make the opening $Dd$ all the larger, we can increase the number of tubes, until they come into contact; and since it is not necessary that they be separated from each other, all these pipes will form a cavity around the walnut $AMF$ in the form of an inverted bell; or else, the machine will be similar to a mold, which is used to cast bells. This cavity will be, therefore, generated by the revolution of the figure $CAMm\ FfDd$ around the axis $OO$, where the circle described by $Cc$ provides an opening at the bottom, and that the [opening] at the top will be a cylindrical surface generated by the revolution of the line $DD$ around the axis.

[Similar to the develops presented so far, in paragraphs LVI, LVII, LVIII, LIX, LX, LXI, LXII, LXIII and LXIV Euler derives the formulation applied to this machine which will not be repeated here for the sake of economy of space and reader's time. Nonetheless, those interested in these detailed formulations can access the paper directly.]

*Description of the Machine*

[Eventually, in paragraphs LXV, LXVI, LXVII, LXVIII, LXIX Euler describes a machine in the form of a parabolic conoid (Fig. 5). The machine will have the shape of a hollow funnel inside, in which the interior surface $AMMDD$ is a parabolic conoid, formed by the revolution of the parabola $AMD$ around the axis $AE$. The exterior surface $ammdd$ is a roughly similar and wider conoid, forming a cavity with the inner surface, through which the water can rise. The water will be admitted into the machine through the bottom opening $cc$ and will be discharged at the top annular opening $DDDD$ to a circular trough $HHHH$, from where it then flows into $P$ through the channel $N$. The machine is free to move around the axis $OO$, which passes through the middle of the funnel and which turns on pivots $R, R$, that are firmly attached at the top and at the bottom. The moment is then applied directly to the axis of the machine, or by means of a wheel; the maneuver and the other parts which will make up the machine can be easily regulated.

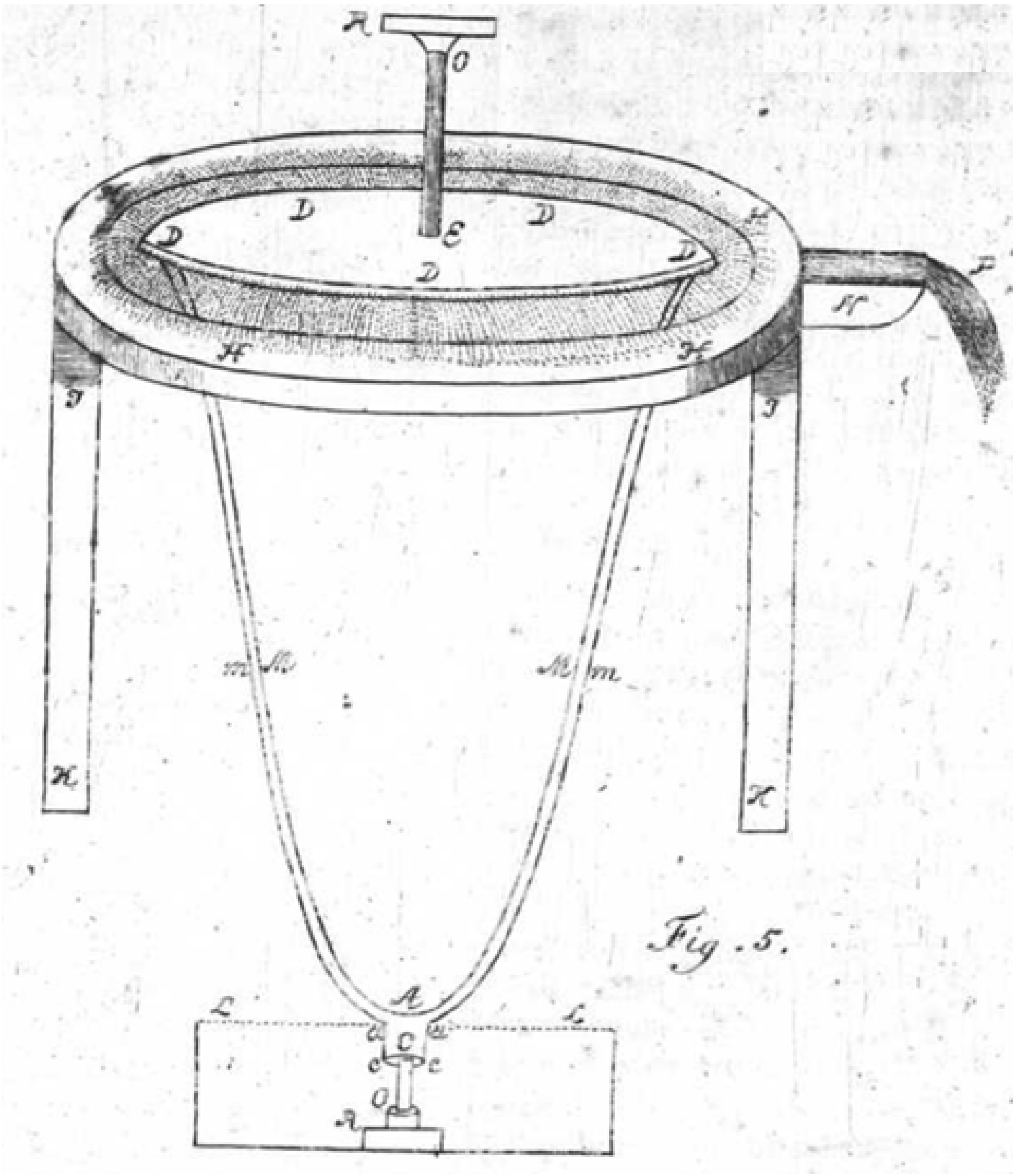

Figure 6: A machine for raising water in the form of a parabolic conoid. (Euler, 1753, Fig. 5)

EXAMPLE

LXX. Set, for example, $AE = 5\frac{5}{8}\,ft$ or $a = l$; and the parabola parameter $g = 3\,ft$, and half the diameter of the above base of the funnel will be $6.846\,ft$, and because of the fold of the upper edge, this half-diameter will rise to $7\,ft$. The diameter $cc$ of the cylinder below will be set around $1\,ft$: the opening of the slot at the top can be set equal to $\frac{1}{50}\,ft$; and the total area of the opening will be $b^2 = 0.88\,ft^2$, or $b^2 = \frac{8}{9}$.

LXXI. Suppose that we wish to apply the force of a man, who acts with a speed of two feet per second, and by taking into consideration that part of this force will be used to overcome friction, then one can set $A = \frac{1}{3}$, and then, we will have that $b^2 = \frac{Ai}{4(\propto+1)a\sqrt{\propto al}} => \frac{8}{9} = \frac{1}{6a^2(\propto+1)a\sqrt{\propto}}$ [this expression was derived in § LX.], or $(\propto+1)\sqrt{\propto} = 0.000768$, from which we have that $\propto = 0.00000059$, hence, we can consider that $\propto = 0$. $D = \frac{Ai}{2a+\frac{A^2i^2}{8a^2b^4l}} \cong \frac{Ai}{2a} = 0.021333\,ft^3$ [this expression was derived in § LXI.]. Then, in one hour a man will be capable of raising to a height of $15\frac{5}{8}$, about $77\,ft^3$ of water.

LXXII. Each revolution of the machine will be achieved during the time of one oscillation of a pendulum of length $q = \frac{2g}{1+\frac{A^2i^2}{16a^3b^4l}} \cong 2g = 6\,ft$ [this expression was derived in § LXI.], or in [$\frac{T}{2} = \pi\sqrt{\frac{q}{2l}} = \pi\sqrt{\frac{6}{2\cdot 15.625}} = 1\frac{1}{3}$ seconds], or the machine makes about 45 turns in one minute. For this purpose, the action of the man is applied to a machine for which $\frac{d^2}{\mu^2} = \frac{i^2q}{8l} = 0.192$ [this expression was derived in § LXII.], and $\frac{d}{\mu} = 0.4381$. Thus, a man can operate directly at the axis of the machine by setting $\mu = 1$, by means of a half a foot crank, applied most conveniently to the axis at the top, allowing the axis to pass through its support $R$ from above, or otherwise, depending on the circumstances.

## 3. Comments

From a borrowed idea of a simple inclined rotating tube with its lower end immersed into water, capable of admitting water at the bottom through an orifice and delivering it through a nozzle at the top, Euler derives the complete formulation of a rotary sprinkler for irrigation purposes. A detailed derivation was accomplished for the calculation of the necessary moment and rotating speed to be applied to the axis of the machine. Next, Euler extends this basic idea by increasing the number of tubes, until they come into contact; and since it is not necessary that they be separated from one another, all these tubes would form a continuous cavity. From these, by the end of the paper, Euler describes a machine in the shape of a funnel, in which the interior surface is a parabolic conoid, with the exterior surface roughly similar and wider, forming a cavity in between, through which the water can rise.

In this publication, Euler uses pioneering approaches by applying the concept of mechanical power to estimate the power needed to drive the machine by the human force, and by measuring time by the length of a pendulum necessary to achieve one oscillation (corresponding to half of the oscillating period). Several design formulas were derived, which were then applied to the funnel type of machine, which was supposed to be operated by one man, by means of a crank, to calculate the discharge at a given height, and for given values of: the parabola parameter, the bottom orifice diameter, and the total area of the discharge opening.

## 4. Conclusions

This is another publication by Euler addressing machines for raising waters, and similar to other writings on the subject, it is also characterized by mathematical rigor associated with practicality, revealing the embodiment of an engineer in a scientist. Here Euler succeeded in generating practical formulas in minute details, which allow the determination of the output of the machine and measures for its construction, supported at the end by a numerical example to show the feasibility of the proposition.